\documentclass[journal]{IEEEtran}
\ifCLASSINFOpdf
\else
\fi
\usepackage{cite}

\usepackage{graphicx}
\usepackage[small]{caption}
\usepackage{subcaption}
\usepackage{color,soul}
\usepackage{array}
\usepackage{amsmath}
\usepackage{booktabs}
\usepackage{footnote}
\makesavenoteenv{tabular}
\makesavenoteenv{table}
\usepackage{color}
\usepackage[linesnumbered,ruled,vlined]{algorithm2e}
\usepackage[noend]{algpseudocode}
\usepackage{setspace}

\begin{document}
%
\title{Hybrid Cell Assignment and Sizing for Power, Area, Delay Product Optimization of SRAM Arrays}
\author{  Ghasem Pasandi,~\IEEEmembership{Student Member,~IEEE,}, Raghav Mehta, Massoud Pedram, \IEEEmembership{Fellow,~IEEE}, and Shahin Nazarian

\thanks{Authors are with the Department of Electrical and Computer Engineering, University of Southern California, Los Angeles,
California, USA. e-mail: pasandi@usc.edu.}

}
\markboth{IEEE Transactions on Circuits and Systems II: Express Brief (DOI: 10.1109/TCSII.2019.2896794)}
{Pasandi \textit{et al.} Hybrid Cell Assignment and Sizing for Power, Area, Delay Product Optimization of SRAM Arrays}
%

\maketitle

\begin{abstract}
Memory accounts for a considerable portion of the total power budget and area of digital systems. Furthermore, it is typically the performance bottleneck of the processing units. Therefore, it is critical to optimize the memory with respect to the product of power, area, and delay (PAD). We propose a hybrid cell assignment method based on multi-sized and dual-$V_{th}$ SRAM cells which improves the PAD cost function by 34\% compared to the conventional cell assignment. We also utilize the sizing of SRAM cells for minimizing the Data Retention Voltage (DRV), and voltages for the read and write operations in the SRAM array. Experimental results in a 32nm technology show that combining the proposed hybrid cell assignment and the cell sizing methods can lower PAD by up to 41\% when compared to the conventional cell design and assignment.
\end{abstract}
\begin{IEEEkeywords}
Data Retention Voltage, DRV, Energy-Efficient, Low-Power, Memory, Reliability, SRAM.
\end{IEEEkeywords}

\IEEEpeerreviewmaketitle

\section{Introduction}
\label{Intro:sec}
\setstretch{0.99}
\IEEEPARstart{L}{ow}-Power circuits and systems have become increasingly popular due to their wide-ranging applications ranging from implantable devices to spacial electronics and mobile devices \cite{Behzad_tvlsi, Darwich_ASurvey, imani2015ultra,shafaei2016energy, Ahmad20171, Gupta_Digital_Computation}. One of the most widely-used and effective techniques to reduce power consumption is to scale down the power supply voltage \cite{Jiao_ISOCC, Gupta_Digital_Computation}. However, conventional designs may fail to operate successfully at low supply voltages, therefore it is necessary to develop new design paradigms. For an SRAM, which consumes a large portion of the energy budget, several designs at the circuit and higher levels have been proposed \cite{Pasandi_TVLSI, Takeda_JSSC, Yi_TCASII, Na_Gong_TCAS, chang2011priority, Pasandi_TED}. In addition, to alleviate short-channel effects (SCE), emerging FinFET and GAA (gate-all-around) device structures and analysis models have been proposed for low voltage operations \cite{cui2016efficient,wang201510nm}.

For applications in which leakage power reduction is the main priority, minimizing the Data Retention Voltage (DRV) is an effective technique to reduce the total leakage power of the standard SRAM. In \cite{qin2004sram}, authors optimized the DRV by choosing suitable values for widths of transistors in the SRAM cell. Another method to decrease the total leakage power is to utilize transistors with different threshold voltages ($V_{th}$). For example, using a combination of transistors with different threshold voltages and oxide thicknesses ($T_{ox}$) is shown to reduce leakage power consumption by up to 40\%\cite{Amelifard_tvlsi}. In such techniques, the only optimization goal is the reduction of the leakage power, whereas for many applications other important metrics such as delay, active power, and area should also be considered.

In this paper, we present a hybrid SRAM cell assignment method that optimizes the product of power, area, and delay (PAD) by assigning multi-sized and dual-$V_{th}$ SRAM cells in the SRAM array. We will discuss and compare six different assignments which have different sizing (normally sized, up-sized I [version 1], and up-sized II [version 2]) with high and low threshold voltages across the array. For normally sized cases, we follow an approach similar to the one discussed in \cite{qin2004sram} to optimize the 6T SRAM cell to achieve the lowest possible DRV subject to a certain noise margin. In an improvement over \cite{qin2004sram}, we calculate an optimal point for DRV by changing both width and length of transistors in the 6T SRAM cell. We also follow a similar design strategy to find the minimum supply voltages for read and write operations. Since these optimization methods will produce different SRAM cell sizing solutions, we compare them using realistic cache design scenarios to select the best design for the SRAM array based on the PAD cost function.
\section{Hybrid SRAM Cell Assignment}
\label {sec:HCS}
\begin{figure}[b]
\centering
\includegraphics[width=0.45\textwidth]{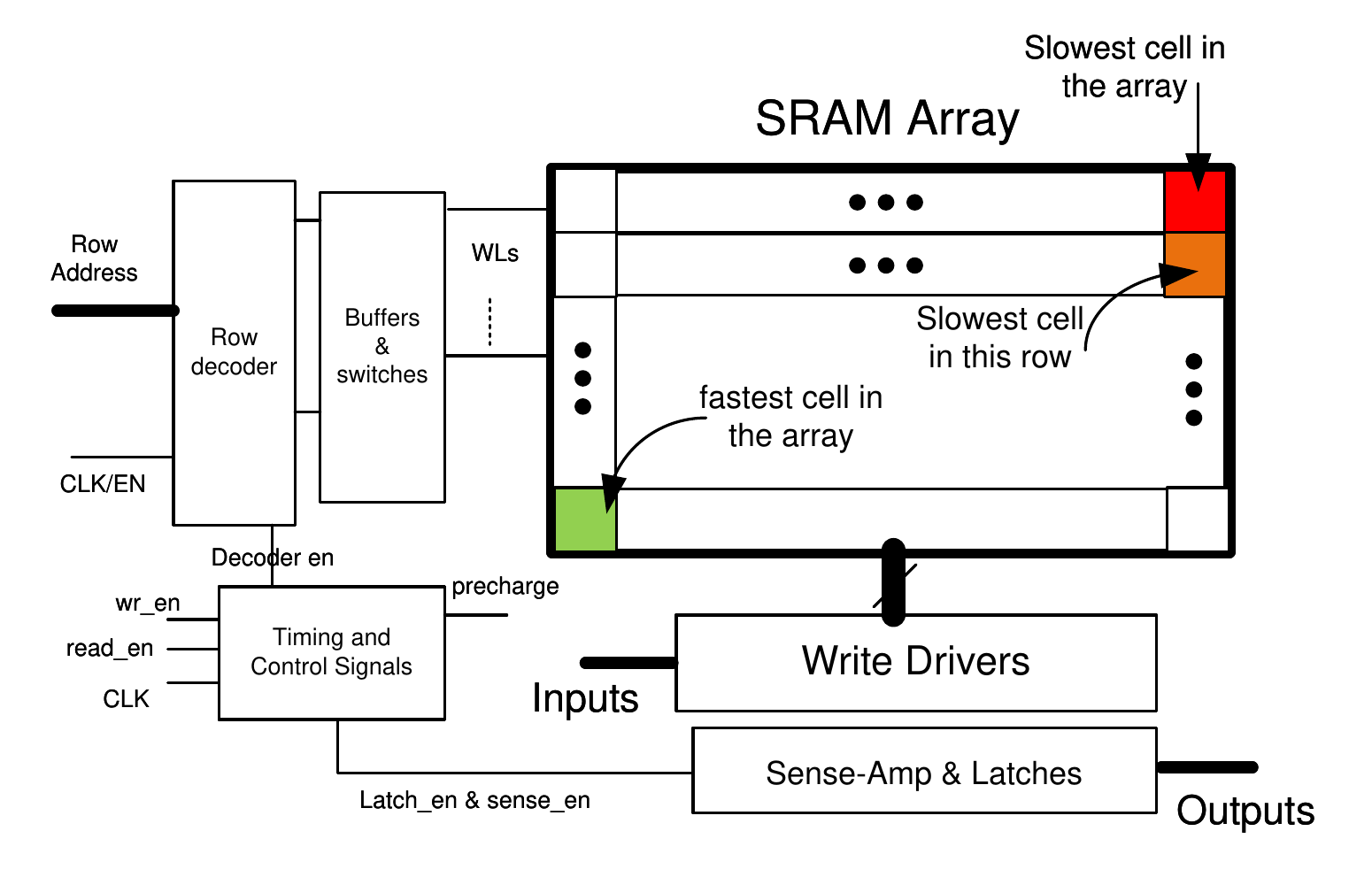}
\caption{Illustration of dependency of the cell placement on the cell read delay.}
\label{fast_slow_cell}
\end{figure}
SRAM cells which are placed farther from the word-line drivers have larger delays as the wordline signal takes longer to reach those cells \cite{Amelifard_tvlsi}. This is depicted in Fig. \ref{fast_slow_cell}. For an SRAM design with a large number of cells in a row, the contribution of wordline wire delay in the overall delay will be significant \cite{shafaei2016energy, shafaei2016minimizing}. For example, for an SRAM with 256 cells per row, operating at 500mV supply voltage in a 32nm technology node \cite{PTM}, the wordline delay is about 35\% of total cell read delay\footnote{delay of wordline and intrinsic delay of an SRAM cell.}. Since the delay of worst case cell will determine the overall delay of the SRAM, we need to reduce this worst case delay.
In the following, we present our multi-sized and multi-$V_{th}$ SRAM cell assignment technique to find the optimal point for the PAD product of the SRAM array.

\subsection{Multi-Sized Cell Assignment}
\label{multi-sub-sec}
Upsizing helps to reduce the read/write operation delays for an SRAM cell. However considering its area and power overheads, we utilize a cost function based on PAD product to account for those overheads. The main task is to find the number of cells that should be sized up to achieve the optimal PAD value for the SRAM array.
\begin{figure}[t]
\centering
\includegraphics[width=0.5\textwidth]{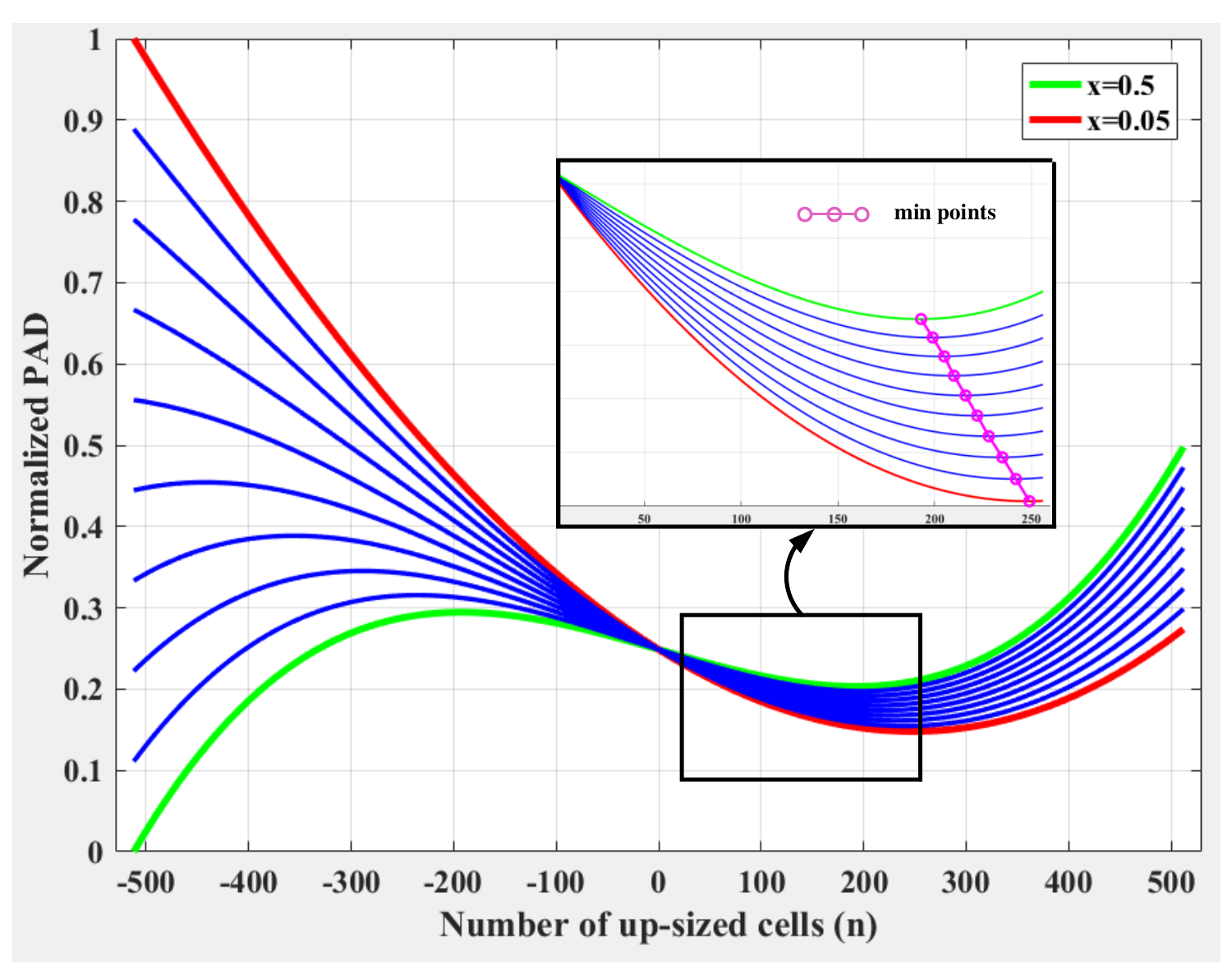}
\caption{PAD cost function versus the number of up-sized SRAM cells in a row of SRAM array. Each graph corresponds to an up-sizing factor, $x$ ($\times W$), ranging from 0.05 to 0.5 with a step of 0.05.}
\label{AD_n}
\end{figure}

Consider a row in the SRAM array with $N$ total number of cells. Our design strategy is to up-size transistors of the last $n$ SRAM cells in the row to make cell read delay of the $N^{th}$ cell less than or equal to the read delay of the cell located in the $(N-n)^{th}$ place. Layout of the up-sized cells are designed such that their height remains the same as normally sized SRAM cells. For this purpose,  width of layout of the up-sized cells are increased by $x$. Eqs. \ref{eq-D_N_n}-\ref{eq-D_N} express the delay of $(N-n)^{th}$ and $N^{th}$ cells. In these equations, delay of the wordline wire is also considered, and wordline is modeled as a distributed RC circuit. We formulate the delay difference between the last cell in a row and the $(N-n)^{th}$ cell as $D_{diff}$ in Eq. \ref{eq-y1}. In our design, we try to minimize the absolute value of $D_{diff}$, keeping in mind that $D_{diff}$ cannot be positive, because the $N^{th}$ cell should not be the critical cell with the highest delay. Note that in Eqs. \ref{eq-D_N_n}-\ref{eq-y1}, $d_1$ stands for the intrinsic delay of the normally sized SRAM cell, and $d_2$ is the intrinsic delay of an up-sized version of the cell. 
Also, $K$ is a technology-dependent constant. $W$ and $H$ are the width and the height of layout of a regular sized 6T SRAM cell, respectively.
{\small
\begin{equation}
Delay^{N-n} = K((N-n)W)^2 + d_1
\label{eq-D_N_n}
\end{equation}
\begin{equation}
Delay^{N} = K(NW+nx)^2 + d_2
\label{eq-D_N}
\end{equation}
\begin{multline}
D_{diff} = Delay^{N} - Delay^{N-n} = \\
(d_2-d_1) + K(n^2x^2 + 2nNWx + (2nN-n^2)W^2)
\label{eq-y1}
\end{multline}
}
The goal of our design is to find the best number of cells that would be up-sized ($n$), and their up-sizing factor $x$. To take the three important factors of SRAMs into account, we used multiplication of power, delay, and area as our optimization cost function Eq. \ref{CF_O}, which is expanded as a function of the SRAM array parameters in Eq. \ref{eq-y2}. Fig. \ref{AD_n} illustrates an example PAD function, as a function of $n$ for different $x$ values. Note that the portion of curves corresponding to negative values for $n$ are also shown to depict the overall trend of PAD versus $n$. 
{\small
\begin{equation}
Cost_{PAD} = Power*Area*Delay
\label{CF_O}
\end{equation}
}
{\small
\begin{multline}
Cost_{PAD} = F(W,H,N,n,x) = \\
p \times (NWH + nxH)(K((N-n)W)^2 + d_1) = \\
\\
p \times KW^2xHn^3 + p \times (-2NKW^2xH+KW^2NWH)n^2 +\\
p \times (xHKW^2N^2 + xHd_1 - NWH \times KW^2 \times 2N)n + \\
p \times (NWH \times KW^2 \times N^2 + NWH \times d_1)\\
\\
\label{eq-y2}
\end{multline}
}
{\small
\begin{equation}
p = \frac{1}{2}C_{load}\frac{V_{DD}^2}{T_{cycle}}\alpha + p_{leakage}
\label{power_eq}
\end{equation}
}
\begin{figure}[b]
\centering
\includegraphics[width=0.43\textwidth]{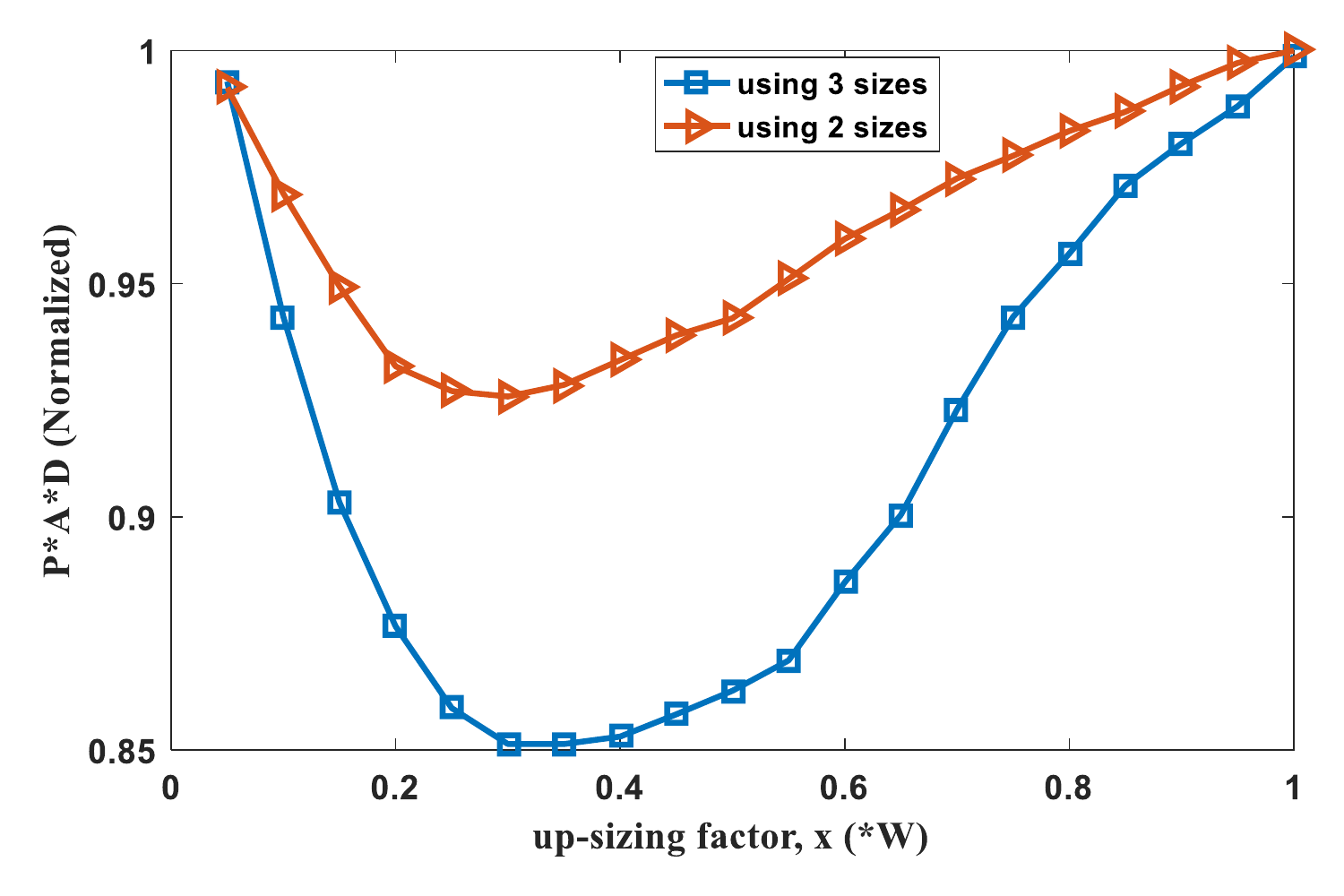}
\caption{Normalized PAD for different sizing choices; in this example, advantage of using three sizing versions over two is clearly seen.}
\label{PAD_n_500mV}
\end{figure}
By having $W=0.80 \mu m$, $H=0.32 \mu m$, $K=0.0014$, $N=256$, and at $V_{DD}=0.5V$, the PAD cost function will be as shown in Fig. \ref{PAD_n_500mV}. As seen in this figure, if we are allowed to use three sizing versions for the SRAM cells (including the normally sized), the improvement on the cost function is much larger. For this case, the improvement is 14\% over the conventional one-sized cell assignment, and for the case of using only two sizing versions, the improvement is 7\%.
\subsection{Multi-$V_{th}$ Cell Assignment}
\label{multi-Vth-subsec}
In this subsection, we extend the procedure in the previous subsection to multi-$V_{th}$ cell assigment using a predictive Bulk-CMOS 32nm Low-Power (LP) and 32nm High-Performance (HP) technologies \cite{PTM}. It is well-known that each additional threshold voltage needs one more mask layer in the fabrication process, which increases the cost and reduces the yield \cite{mukhopadhyay2005modeling}. Therefore, it is common to limit the multi-$V_{th}$ cell libraries to dual-$V_{th}$, i.e., high-$V_{th}$ and low-$V_{th}$ transistors. Dual-$V_{th}$ assignment is a well-known optimization technique, e.g., a dual-$V_{th}$, dual-$T_{ox}$ solution was proposed in \cite{Amelifard_tvlsi} to reduce the overall leakage power consumption. However, in this paper, we incorporate dual-$V_{th}$ assignment to our multi-sizing algorithm, discussed in the previous section, to minimize the PAD product. For the simplicity of the fabrication process, we assume all the transistors in an SRAM cell are chosen to have the same threshold voltage, i.e., either low-$V_{th}$ or high-$V_{th}$. 
\subsection{Multi-Sized Dual-$V_{th}$ Cell Assignment}
\label{Multi-size-Dual-Vth}
\begin{figure}[t]
\centering
\includegraphics[width=0.5\textwidth]{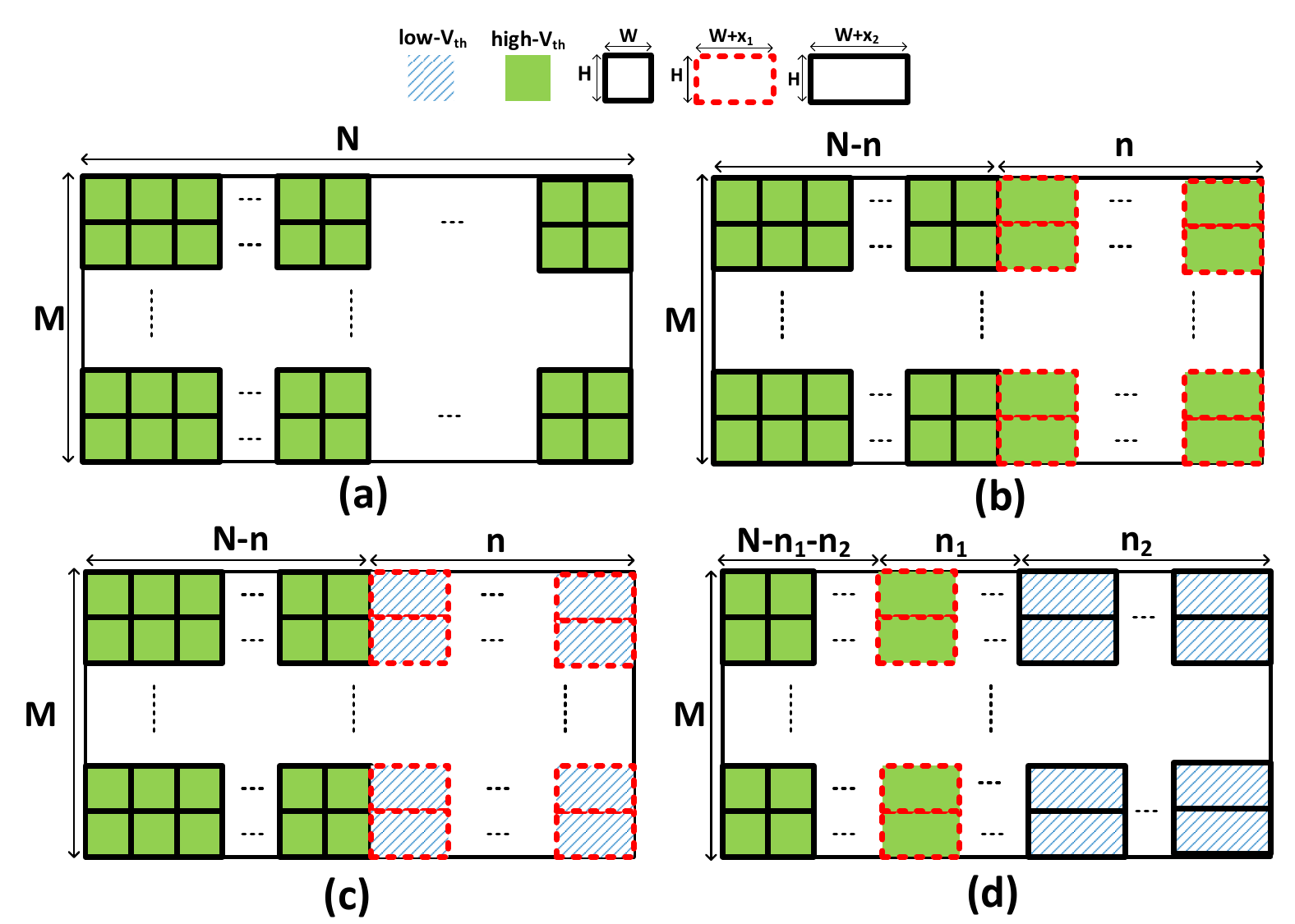}
\caption{Four different cell assignments in the SRAM array corresponding to the following triplets defined in Section \ref{Multi-size-Dual-Vth}; (a) ($1_H$,0,0), (b) ($1_H$,$2_H$,0), (c) ($1_H$,$2_L$,0), (d) ($1_H$,$2_H$,$3_L$).}
\label{Placement}
\end{figure}
Considering both multi-sized and dual-$V_{th}$ assignments, we develop a hybrid cell assignment in the SRAM array. The following six different cell assignments are considered. Fig. \ref{Placement} shows four of these cases.\\
\textbf{1.} All cells are high-$V_{th}$ and normally sized. \\
\textbf{2.} All cells are high-$V_{th}$, among which $N$-$n$ cells are normally sized, and the rest ($n$ cells) are up-sized I [version 1].\\
\textbf{3.} All cells are high-$V_{th}$, among which $N$-$n_1$-$n_2$ cells are normally sized, $n_1$ cells are up-sized I [version 1], and the rest of the cells are up-sized II [version 2].\\
\textbf{4.} All cells are normally sized, among which $N$-$n$ cells are high-$V_{th}$, and the rest of them are low-$V_{th}$.\\
\textbf{5.} $N$-$n$ cells are high-$V_{th}$ and they are normally sized. The rest of the cells are low-$V_{th}$ and are up-sized I [version 1].\\
\textbf{6.} $N$-$n_1$-$n_2$ cells are high-$V_{th}$ and they are normally sized, $n_1$ cells are low-$V_{th}$ and are up-sized I [version 1], and the rest of them are low-$V_{th}$ and are up-sized II [version 2].

Each design of different configuration is represented by a triplet \textit{($p_{a}$,$q_{b}$,$r_{c}$)} where the first entry, p, corresponds to the first $(N$-$n_1$-$n_2)$ cells in the SRAM array; the second entry, q, corresponds to the next $(n_1)$ cells, and the third entry, r, corresponds to the last $(n_2)$ cells. Each entry is either zero, one, two or three, if the corresponding cells are not used, are normally sized, are up-sized I (up-sized by $x_1$ amount), and are up-sized II (up-sized by $x_2$ amount), respectively. The subscript corresponds to low-$V_{th}$ or high-$V_{th}$ by having a letter $L$ or $H$, respectively. For example, ($1_{H}$,0,0) corresponds to the original configuration where all $N$ cells are normally-sized and with high-$V_{th}$, and ($1_{H}$,$2_{H}$,$3_{L}$) corresponds to a configuration with $(N$-$n_1$-$n_2)$ first cells with nominal sizing and high-$V_{th}$, up-sized I for the next $n_1$ cells with high-$V_{th}$, and the last $n_2$ cells with up-sized II and low-$V_{th}$. It is clear that a configuration with (0,0,0) does not exist. 
\begin{algorithm} [t]
{\small
\caption{Hybrid SRAM Cell Assignment}
\DontPrintSemicolon 
\KwIn{$N$: Number of cells in a row, \\$\mathcal{V}$: Set of allowed threshold voltages, \\$\#Sizes$: Number of allowed size versions, \\$\mathcal{T}$: Technology }
\KwOut{Best cell assignment}
\tcp{\small{Initializing parameters:}} 
\label{place_algorithm}
Set $V_{DD}$, length and width of transistors, H (height) and W (width) of SRAM cell's layout;\\
\tcp{\small{Extracting intrinsic delay for normally} sized cell ($d_1$):}
resp = system(hspice -i input.sp -o input);\\
$d_1$ = Extract\_delay(input.mt0); \\
\tcp{Performing cell assignment:}
left\_pointer = 1; \\
\For{iterator $i$ in range($\#Sizes-1$)}{
	\For {up-sizing factor $x \in$ [0:0.05:1]$\times W$ and $V_{th} \in \mathcal{V}$}{
    	\small{Find $n$, optimum number of right most cells to be up-sized, and their best $V_{th}$ assignments;}
	}
	Save the best cell assignment up to now;\\
	left\_pointer = $N-n+1$;\\
	$N=n$;
}

\Return{The best obtained cell assignment;}\;
}
\end{algorithm}

Algorithm \ref{place_algorithm} shows the pseudocode of our hybrid cell assignment approach. After initializing some parameters in line 1, intrinsic delay of a normally sized SRAM cell with high-$V_{th}$ is extracted in lines 2-3. The best cell assignment for a row in the SRAM array is then found in lines 4-10. More specifically, in the $for$ loop shown in lines 6-7, the best threshold voltages together with the best sizing for $1^{th}$ to $N^{th}$ cells are found. At the end of this $for$ loop, number of right most cells ($n$) that should be up-sized and their threshold voltages are found. In the next iteration, this loop is run on the $N$-$n$+$1^{th}$ cell to the $N^{th}$ cell, and the length of the row is set to $n$. The procedure repeats $\#Sizes$ times, and finds the final cell assignment for the entire row. This cell assignment will be used for other rows as well. 

Tables \ref{tab:cell-assignment}-\ref{tab:cell-assignment_90nm} show the set of configurations along with their improvements on the overall PAD cost function compared with the conventional cell assignment. As seen in Table \ref{tab:cell-assignment}, having $N$=$256$, if we are allowed to use three sizing versions for the SRAM cells in an array with $N$-$n_1$-$n_2$=$68$ regular sized cells , $n_1$=$70$ up-sized I with high-$V_{th}$ cells and $n_2$=$118$ up-sized II cells with low-$V_{th}$ cells, the improvement on the cost function is much higher. For this case, the improvement is 34\% over the conventional one-size high-$V_{th}$ assignment, and for the case of using only two sizing versions for the SRAM cells, the improvement is 16\%. By considering a 10\% variation in the threshold voltage and sizes of cells (modeling the process variation), the 34\% improvement for ($1_H$,$2_H$,$3_L$) cell assignment will be decreased to 10\%.
\begin{table}[t]
  \centering
  \scriptsize
  \caption{Amounts of reduction in PAD cost function for different cell assignments in 32nm technology.}
    \begin{tabular}{ccc}
    \toprule
    Cell Assignment & Cell Counts & Cost Reduction(\%) \\
    \midrule
    ($1_H$, 0  , 0  ) & (256, 0, 0) & - \\
    \midrule
    ($1_H$, $2_H$, 0  ) & (121, 135, 0) & 7 \\
    \midrule
    ($1_H$, $2_H$, $3_H$) & (70, 74, 112) & 14 \\
    \midrule
    ($1_H$, $1_L$, 0 ) & (124, 132, 0) & 4 \\
    \midrule
    ($1_H$, $2_L$, 0 ) & (119, 137, 0) & 16 \\
    \midrule
    ($1_H$, $2_H$, $3_L$) & (68, 70, 118) & 34 \\
    \bottomrule
    \end{tabular}%
  \label{tab:cell-assignment}%
\end{table}%
\begin{table}[t]
  \centering
  \scriptsize
  \caption{Amounts of reduction in PAD cost function for different cell assignments in 90nm technology.}
    \begin{tabular}{ccc}
    \toprule
    Cell Assignment & Cell Counts & Cost Reduction(\%) \\
    \midrule
    ($1_H$, 0  , 0  ) & (256, 0, 0) & - \\
    \midrule
    ($1_H$, $2_H$, 0  ) & (228, 28, 0) & 6 \\
    \midrule
    ($1_H$, $2_H$, $3_H$) & (189, 65, 2) & 12 \\
    \midrule
    ($1_H$, $1_L$, 0 ) & (150, 106, 0) & 27 \\
    \midrule
    ($1_H$, $2_L$, 0 ) & (145, 111, 0) & 28 \\
    \midrule
    ($1_H$, $2_H$, $3_L$) & (13, 147, 96) & 40 \\
    \bottomrule
    \end{tabular}%
  \label{tab:cell-assignment_90nm}%
\end{table}%

Please note that in the case of driving SRAM cells from two sides of the SRAM array, we can use the above optimization/design procedure for $N'$=$N/2$ to find the best cell assignment for the first half of the array. The other half will be the mirror of the first one. Note that in this case, some portions (i.e., the cells in the middle) of the SRAM array will end up with higher size and/or lower threshold voltages. 

Our hybrid SRAM cell assignment algorithm is applicable to various devices and technologies including standard Bulk-CMOS, FinFET, and FDSOI. However, the optimal cell assignment depends on the device type and technology node. More precisely, the number of upsized cells or cells with higher threshold voltage values may be different for FinFETs and FDSOIs, and different in 32nm technology when compared to 14nm technology. This also means that the PAD improvement varies from one technology node or device type to another.
\subsection{Reliable SRAM Cell Design}
\label{sec:DRV}
\begin{figure}[t]
\centering
\includegraphics[width=0.35\textwidth]{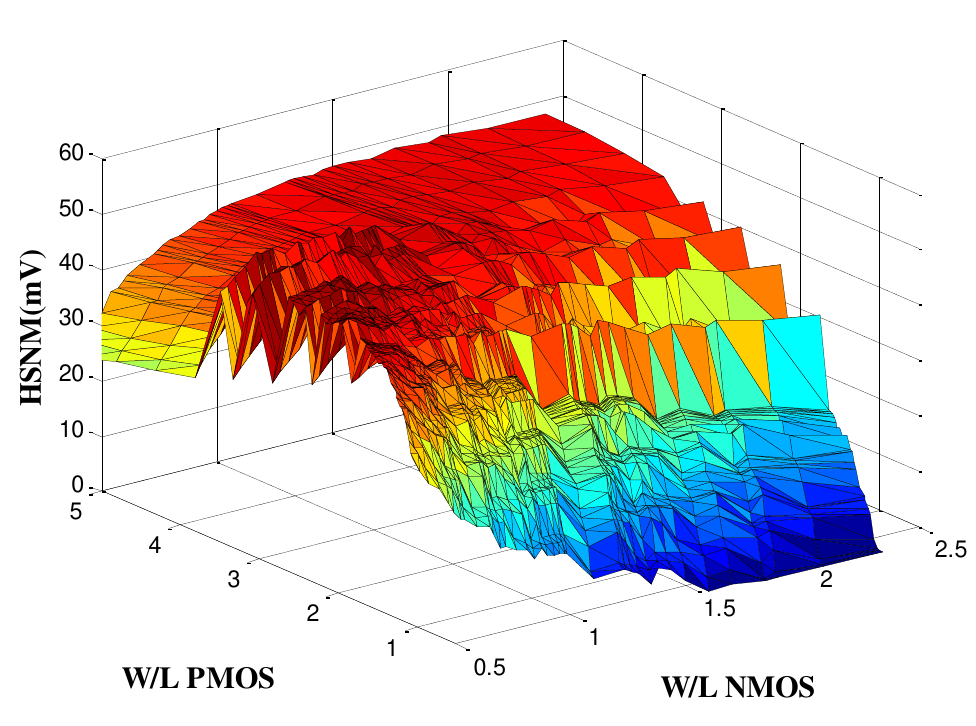}
\caption{Hold Noise Margin as a function of both NMOS and PMOS transistors' widths and lengths.}
\label{HSNM_NMOS_P}
\end{figure}

In \cite{qin2004sram}, authors have formulated the DRV of a 6T SRAM cell based on sizes of transistors and some technology parameters for a 0.13$\mu$m industrial technology. Using this formula, the DRV value for a predictive 32nm Bulk-CMOS technology (PTM) \cite{PTM} is calculated as 11mV, which is smaller than the thermal noise (26mV). Using 26mV as a starting voltage and considering the variation on the threshold voltage, the final DRV after adding a 100mV guard band voltage to account for larger memories will be 194mV. Fig. \ref{HSNM_NMOS_P} shows the Hold Static Noise Margin (HSNM) for joint sweeping of NMOS and PMOS transistors' width and length values. The best design has the SNM value of 59mV, that we shall set as a minimum required SNM for designing other SRAM cells.
By following the similar design methodologies for minimizing the supply voltages for read and write operations, new designs (sizes of transistors) will be achieved. In Section \ref{sec:sim}, we provide the results for these design methodologies.
\section{Simulation Results}
\label{sec:sim}
\begin{figure}[t]
        \centering
        \begin{subfigure}[!t]{0.21\textwidth}
                \centering
                \includegraphics[width=\textwidth]{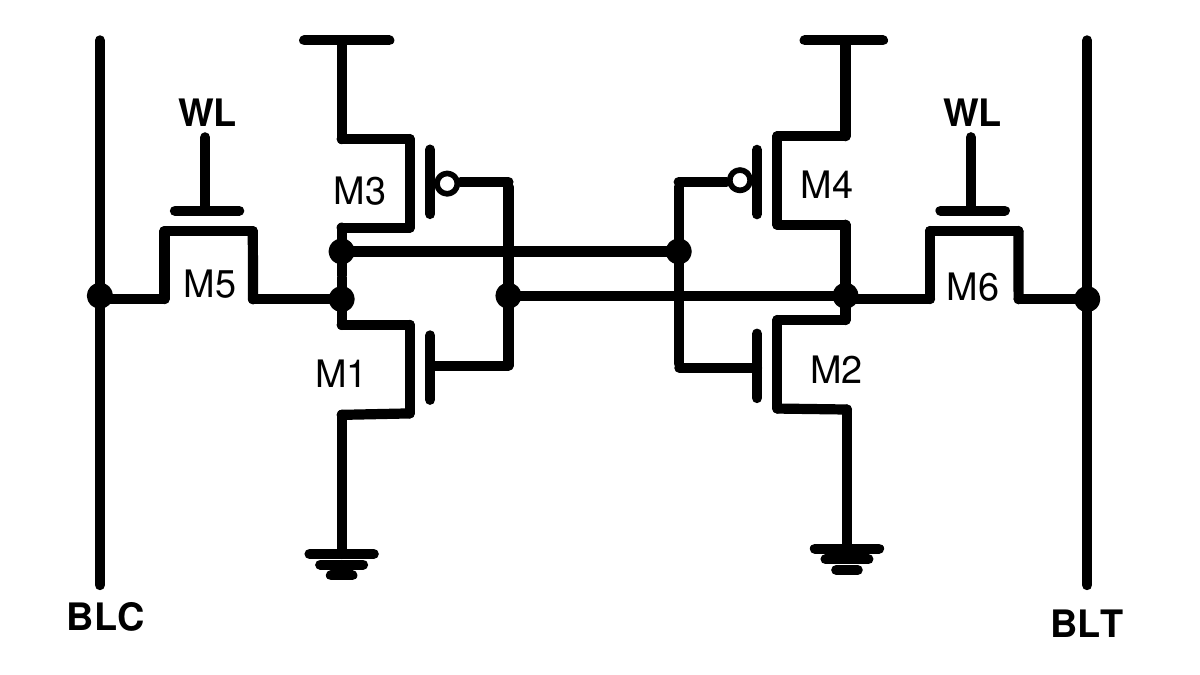}
                \caption{}
                \label{6T_cell}
        \end{subfigure}
        \begin{subfigure}[!t]{0.23\textwidth}
                \centering
                \includegraphics[width=\textwidth]{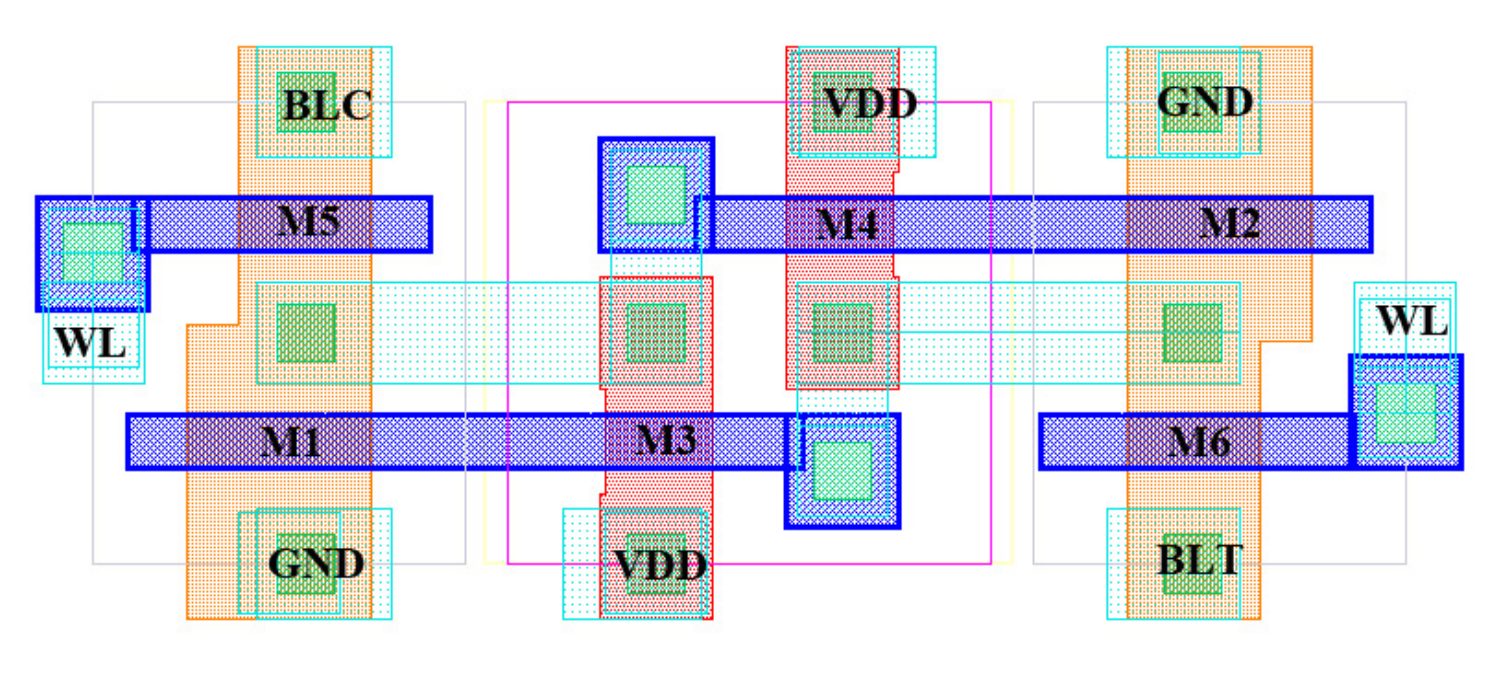}
                \caption{}
                \label{6T_Layout}
        \end{subfigure}
        \caption{Conventional 6T SRAM cell, (a) transistor-level schematic, (b) layout, showing the name of transistors that are used in this paper. The layout is for DRV-based sizing mentioned in Table \ref{tab:sizing-read-write}.}
        \label{6T-SRAM}
\end{figure}
\begin{figure}[t]
        \centering
        \begin{subfigure}[!t]{0.45\textwidth}
                \centering
                \includegraphics[width=\textwidth]{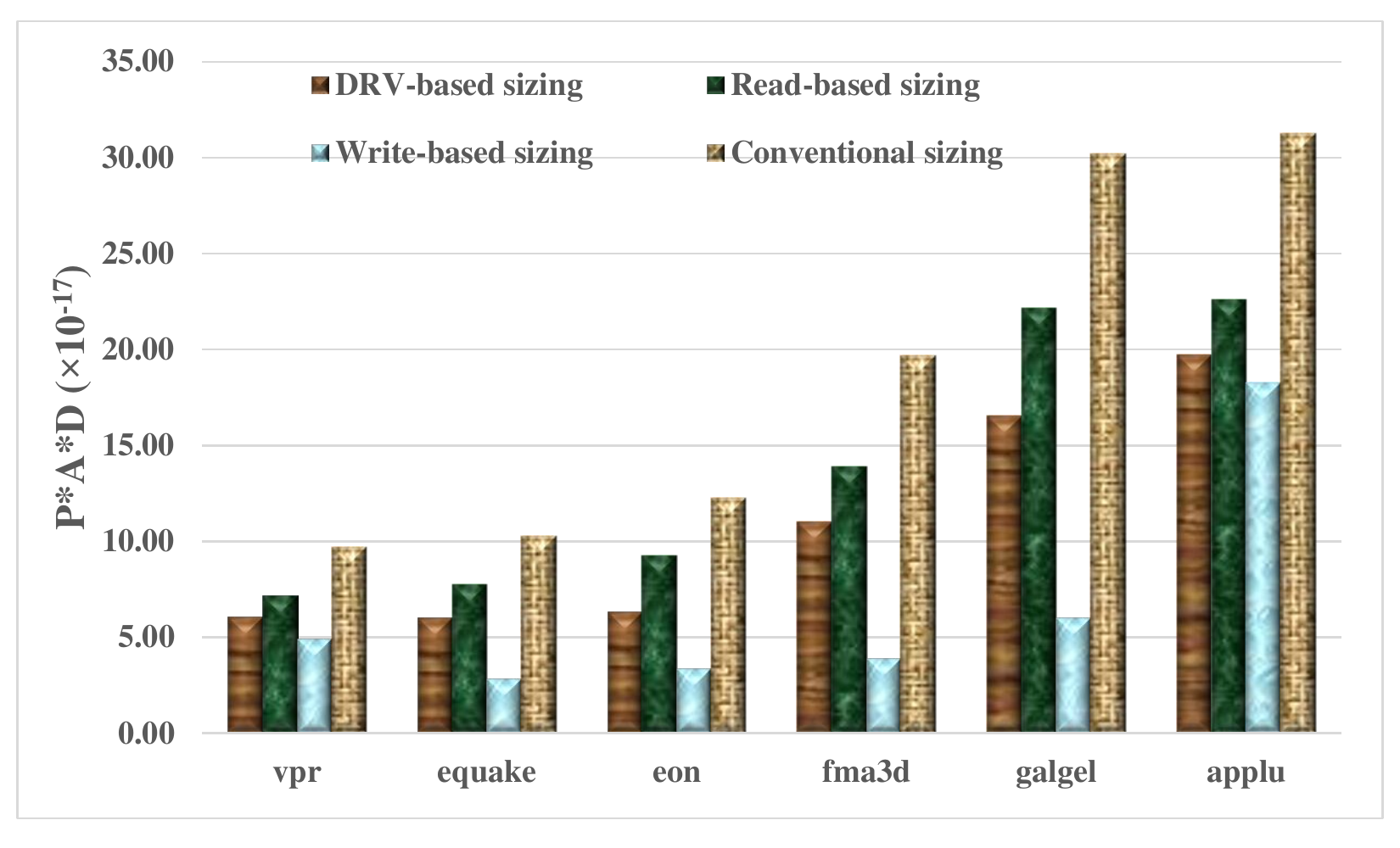}
                \caption{}
                \label{energy_benchmark2}
        \end{subfigure}
        \begin{subfigure}[!t]{0.45\textwidth}
                \centering
                \includegraphics[width=\textwidth]{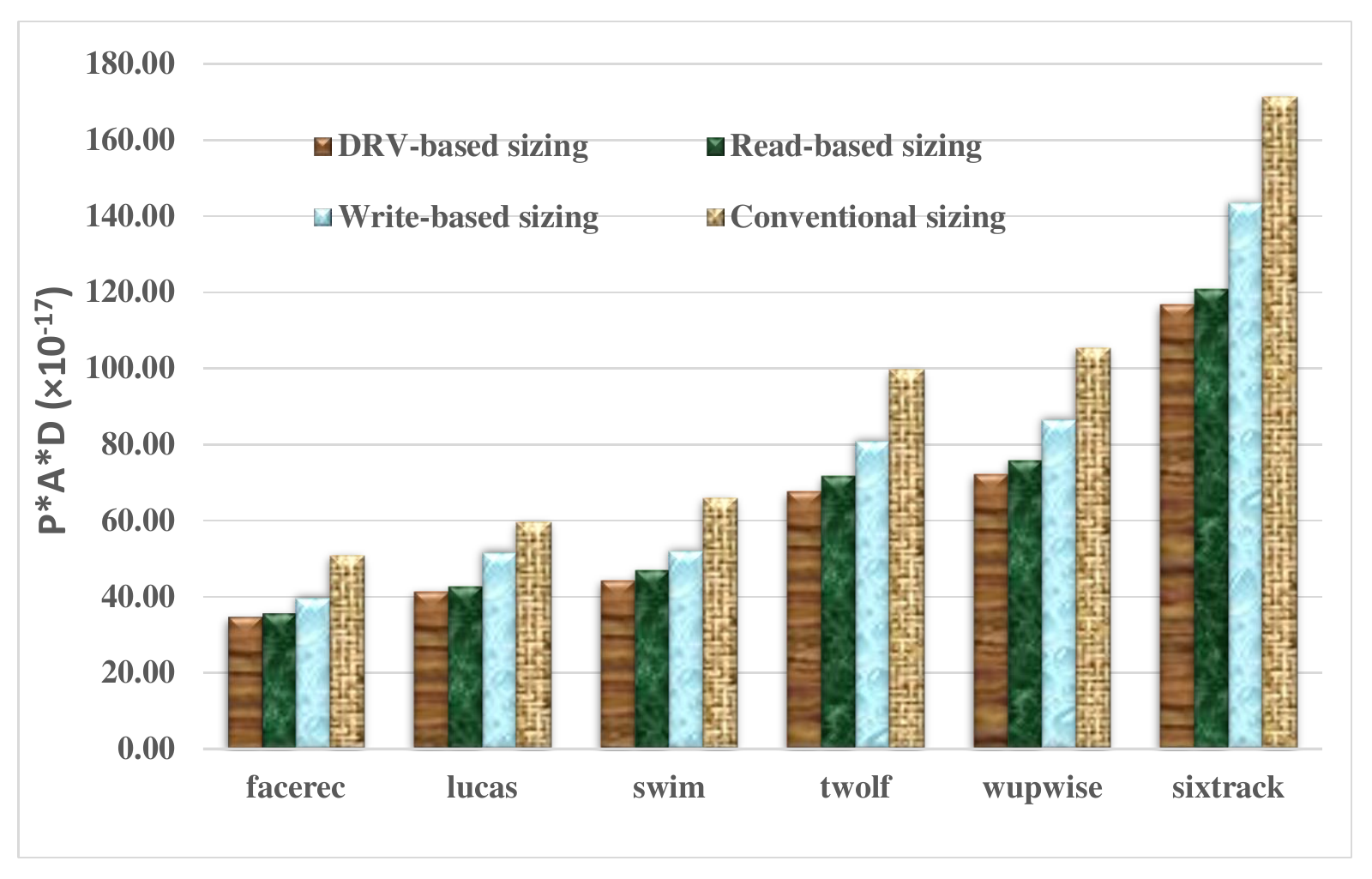}
                \caption{}
                \label{energy_benchmark}
        \end{subfigure}
        \caption{Comparing PAD product cost function for different benchmarks in four design methods, benchmarks with (a) small idle times (hot caches), and (b) large idle times (cold caches). }
        \label{energy_benchmark_}
\end{figure}

\begin{table}[t]
\scriptsize 
  \centering
  \caption{Values for widths and lengths of transistors in optimization for hold (DRV), read, and write operations (the values are multiples of $L_{min}$ in the used technology).}
    \begin{tabular}{ccccccc}
    \toprule
     Design Method  & \multicolumn{2}{c}{DRV-based}   & \multicolumn{2}{c}{Read-based} & \multicolumn{2}{c}{Write-based} \\
     \midrule
     technology & 32nm & 90nm & 32nm & 90nm & 32nm & 90nm  \\
    \midrule
    W M1/M2 & 4.5 &  2.0 & 3.5&  3.0   & 3.5& 3.5   \\
    \midrule
    W M3/M4 & 2.5&  2.5 & 4.0&   2.5  & 2.0& 2.0  \\
    \midrule
    W M5/M6 & 2.5&  1.5  & 3.0&   1.5  & 5.0&  3.0   \\
    \midrule
    L M1/M2 & 2.0& 3.0 & 1.5&   3.0  & 1.5& 1.5  \\
    \midrule
    L M3/M4 & 3.0& 3.0 & 3.5&  2.0  & 4.0& 2.0   \\
    \midrule
    L M5/M6 & 1.0& 2.0  & 3.0&  4.0  & 1.0& 2.0   \\
    \bottomrule
    \end{tabular}%
  \label{tab:sizing-read-write}%
\end{table}%
We designed and optimized 6T SRAM cell for three different approaches (DRV-based, read-based, and write-based sizing).
Table \ref{tab:sizing-read-write} shows the final values for sizes of transistors for each of these methods. We also considered the conventional cell sizing method.
To compare these four cell sizing methods in a real set-up, we designed four 32kb SRAMs (each with a single block), in each of them the base cell is chosen from one of the four mentioned cell sizing methods. We applied our best hybrid cell assignment technique to all of these four memories. PAD product cost function was used to compare different designs. Fig. \ref{energy_benchmark2} shows PAD for running couple of benchmarks with small idle times (hot caches), and Fig. \ref{energy_benchmark} depicts the results for benchmarks with large idle times (cold caches). All benchmarks are from SPEC CPU2000 \cite{SPEC2000}. We used CACTI for extracting these results. Also, Hspice 2016 and Matlab 2016 are used for SRAM characterizations. As seen, write-based method with hybrid cell assignment works better for hot caches and shows about 41\% improvements on the cost function over the conventional sizing for \textit{applu} benchmark. For cold caches, the DRV-based method is better which shows 32\% improvement over the conventional sizing in \textit{sixtrack} benchmark. Thus, our recommendation is to use write-based method for hot caches such as L1 instruction cache, and DRV-based for cold caches such as L2 cache, and apply our hybrid cell assignment to all of these caches.

\section{Conclusions}
\label{cons:sec}
In this paper, we proposed a hybrid cell assignment for SRAM, which is based on using multi-sized dual-$V_{th}$ transistors in the SRAM array. In addition, a DRV-based optimization design method for cell sizing is presented. In this method, width and length of transistors in the 6T SRAM cell are optimized to achieve the smallest DRV subject to a minimum noise margin. Using this method for optimizing read and write operations, two new designs are obtained. Simulation results for SPEC CPU2000 benchmarks confirmed significant reduction in PAD cost function for both hot caches such as L1 and cold caches such as L2 caches.

\ifCLASSOPTIONcaptionsoff
  \newpage
\fi
\bibliographystyle{IEEEtran}

\begin{thebibliography}{10}
\providecommand{\url}[1]{#1}
\csname url@samestyle\endcsname
\providecommand{\newblock}{\relax}
\providecommand{\bibinfo}[2]{#2}
\providecommand{\BIBentrySTDinterwordspacing}{\spaceskip=0pt\relax}
\providecommand{\BIBentryALTinterwordstretchfactor}{4}
\providecommand{\BIBentryALTinterwordspacing}{\spaceskip=\fontdimen2\font plus
\BIBentryALTinterwordstretchfactor\fontdimen3\font minus
  \fontdimen4\font\relax}
\providecommand{\BIBforeignlanguage}[2]{{%
\expandafter\ifx\csname l@#1\endcsname\relax
\typeout{** WARNING: IEEEtran.bst: No hyphenation pattern has been}%
\typeout{** loaded for the language `#1'. Using the pattern for}%
\typeout{** the default language instead.}%
\else
\language=\csname l@#1\endcsname
\fi
#2}}
\providecommand{\BIBdecl}{\relax}
\BIBdecl

\bibitem{Behzad_tvlsi}
B.~Ebrahimi, M.~Rostami, A.~Afzali-Kusha, and M.~Pedram, ``Statistical design
  optimization of finfet sram using back-gate voltage,'' \emph{IEEE
  Transactions on Very Large Scale Integration (VLSI) Systems}, vol.~19,
  no.~10, pp. 1911--1916, Oct 2011.

\bibitem{Darwich_ASurvey}
M.~Darwich, A.~Abdelgawadf, and M.~Bayoumi, ``A survey on the power and
  robustness of finfet sram,'' in \emph{2016 IEEE 59th International Midwest
  Symposium on Circuits and Systems (MWSCAS)}, Oct 2016, pp. 1--4.

\bibitem{imani2015ultra}
M.~Imani, M.~Jafari, B.~Ebrahimi, and T.~S. Rosing, ``Ultra-low power finfet
  based sram cell employing sharing current concept,'' \emph{Microelectronics
  Reliability, Available online}, vol.~10, 2015.

\bibitem{shafaei2016energy}
A.~Shafaei and M.~Pedram, ``Energy-efficient cache memories using a dual-vt 4t
  sram cell with read-assist techniques,'' in \emph{2016 Design, Automation \&
  Test in Europe Conference \& Exhibition (DATE)}.\hskip 1em plus 0.5em minus
  0.4em\relax IEEE, 2016, pp. 457--462.

\bibitem{Ahmad20171}
\BIBentryALTinterwordspacing
S.~Ahmad, M.~K. Gupta, N.~Alam, and M.~Hasan, ``Low leakage single bitline
  \uppercase{9T (SB9T)} static random access memory,'' \emph{Microelectronics
  Journal}, vol.~62, pp. 1 -- 11, 2017. [Online]. Available:
  \url{http://www.sciencedirect.com/science/article/pii/S0026269216300945}
\BIBentrySTDinterwordspacing

\bibitem{Gupta_Digital_Computation}
S.~Gupta, A.~Raychowdhury, and K.~Roy, ``Digital computation in subthreshold
  region for ultralow-power operation: A device--circuit--architecture codesign
  perspective,'' \emph{Proceedings of the IEEE}, vol.~98, no.~2, pp. 160--190,
  2010.

\bibitem{Jiao_ISOCC}
H.~Jiao and V.~Kursun, ``Power gated sram circuits with data retention
  capability and high immunity to noise: A comparison for reliability in low
  leakage sleep mode,'' in \emph{SoC Design Conference (ISOCC), 2010
  International}, Nov 2010, pp. 5--8.

\bibitem{Pasandi_TVLSI}
G.~Pasandi and S.~M. Fakhraie, ``A 256-kb 9\uppercase{T} near-threshold
  \uppercase{SRAM} with 1k cells per bitline and enhanced write and read
  operations,'' \emph{IEEE Transactions on Very Large Scale Integration (VLSI)
  Systems}, vol.~23, no.~11, pp. 2438--2446, Nov 2015.

\bibitem{Takeda_JSSC}
K.~Takeda, Y.~Hagihara, Y.~Aimoto, M.~Nomura, Y.~Nakazawa, T.~Ishii, and
  H.~Kobatake, ``A read-static-noise-margin-free \uppercase{SRAM} cell for
  low-vdd and high-speed applications,'' \emph{IEEE Journal of Solid-State
  Circuits}, vol.~41, no.~1, pp. 113--121, 2006.

\bibitem{Yi_TCASII}
Y.-W. Chiu, Y.-H. Hu, M.-H. Tu, J.-K. Zhao, Y.-H. Chu, S.-J. Jou, and C.-T.
  Chuang, ``40 nm bit-interleaving 12\uppercase{T} subthreshold
  \uppercase{SRAM} with data-aware write-assist,'' \emph{IEEE Transactions on
  Circuits and Systems I: Regular Papers}, vol.~61, no.~9, pp. 2578--2585, Sept
  2014.

\bibitem{Na_Gong_TCAS}
N.~Gong, S.~Jiang, A.~Challapalli, S.~Fernandes, and R.~Sridhar, ``Ultra-low
  voltage split-data-aware embedded \uppercase{SRAM} for mobile video
  applications,'' \emph{IEEE Transactions on Circuits and Systems II: Express
  Briefs}, vol.~59, no.~12, pp. 883--887, Dec 2012.

\bibitem{chang2011priority}
I.~J. Chang, D.~Mohapatra, and K.~Roy, ``A priority-based \uppercase{6T/8T}
  hybrid \uppercase{SRAM} architecture for aggressive voltage scaling in video
  applications,'' \emph{IEEE Transactions on Circuits and Systems for Video
  Technology}, vol.~21, no.~2, pp. 101--112, 2011.

\bibitem{Pasandi_TED}
G.~Pasandi and S.~M. Fakhraie, ``An 8\uppercase{T} low-voltage and low-leakage
  half-selection disturb-free \uppercase{SRAM} using
  \uppercase{B}ulk-\uppercase{CMOS} and \uppercase{F}in\uppercase{FET}s,''
  \emph{IEEE Transactions on Electron Devices}, vol.~61, no.~7, pp. 2357--2363,
  July 2014.

\bibitem{cui2016efficient}
T.~Cui, J.~Li, A.~Shafaei, S.~Nazarian, and M.~Pedram, ``An efficient timing
  analysis model for 6t finfet sram using current-based method.'' in
  \emph{ISQED}, 2016, pp. 263--268.

\bibitem{wang201510nm}
L.~Wang, A.~Shafaei, S.~Chen, Y.~Wang, S.~Nazarian, and M.~Pedram, ``10nm
  gate-length junctionless gate-all-around (jl-gaa) fets based 8t sram design
  under process variation using a cross-layer simulation,'' in
  \emph{SOI-3D-Subthreshold Microelectronics Technology Unified Conference
  (S3S), 2015 IEEE}.\hskip 1em plus 0.5em minus 0.4em\relax IEEE, 2015, pp.
  1--2.

\bibitem{qin2004sram}
H.~Qin, Y.~Cao, D.~Markovic, A.~Vladimirescu, and J.~Rabaey, ``Sram leakage
  suppression by minimizing standby supply voltage,'' in \emph{Quality
  Electronic Design, 2004. Proceedings. 5th International Symposium on}.\hskip
  1em plus 0.5em minus 0.4em\relax IEEE, 2004, pp. 55--60.

\bibitem{Amelifard_tvlsi}
B.~Amelifard, F.~Fallah, and M.~Pedram, ``Leakage minimization of sram cells in
  a dual- $v_t$ and dual- $t_{\rm ox}$ technology,'' \emph{IEEE Transactions on
  Very Large Scale Integration (VLSI) Systems}, vol.~16, no.~7, pp. 851--860,
  July 2008.

\bibitem{shafaei2016minimizing}
A.~Shafaei, H.~Afzali-Kusha, and M.~Pedram, ``Minimizing the energy-delay
  product of sram arrays using a device-circuit-architecture co-optimization
  framework,'' in \emph{Proceedings of the 53rd Annual Design Automation
  Conference}.\hskip 1em plus 0.5em minus 0.4em\relax ACM, 2016, p. 107.

\bibitem{PTM}
\BIBentryALTinterwordspacing
A.~S. University. (2013) Predictive technology model (ptm). [Online].
  Available: \url{http://ptm.asu.edu/}
\BIBentrySTDinterwordspacing

\bibitem{mukhopadhyay2005modeling}
S.~Mukhopadhyay, H.~Mahmoodi, and K.~Roy, ``Modeling of failure probability and
  statistical design of sram array for yield enhancement in nanoscaled cmos,''
  \emph{IEEE transactions on computer-aided design of integrated circuits and
  systems}, vol.~24, no.~12, pp. 1859--1880, 2005.

\bibitem{SPEC2000}
\BIBentryALTinterwordspacing
J.~F. Cantin and M.~D. Hill. (2003) SPEC CPU2000 benchmarks. [Online].
  Available:
  \url{http://research.cs.wisc.edu/multifacet/misc/spec2000cache-data/}
\BIBentrySTDinterwordspacing

\end{thebibliography}

\end{document}